
\magnification=\magstep1
%
\font\titlefont=cmr10 at 18pt

\font\namefont=cmti10 at 16pt
\font\headfont=cmbx10 at 12pt
\font\subheadfont=cmbx10 at 10pt
\font\subsubheadfont=cmti10

%
%
\newcount\Headernumber \Headernumber=0
%
%
\def\sectskip{\vskip 10pt plus 0pt minus 6pt}
\def\medsectskip{\vskip 5pt plus 0pt minus 3pt}
\def\smallsectskip{\vskip 3pt plus 0pt minus 1pt}
%
%
\def\section#1{\par\sectskip\nobreak%
   \noindent{\headfont
#1}\smallsectskip\nobreak\noindent\ignorespaces}
\def\subsection#1{\par\medsectskip\nobreak%
   \noindent{\subheadfont #1}%
       \smallsectskip\noindent\ignorespaces}
\def\subsubsection#1{\par\smallsectskip \nobreak%
  \noindent{\subsubheadfont #1}
      \smallsectskip\nobreak\noindent\ignorespaces}
\def\headline{\ifnum\Headernumber=0{}\else\ifodd\pageno\tenit%
  \title\hfil\tenrm\folio
   \else\tenrm\folio\hfil\tenit\author\rm\fi\fi}
%
%
%
\def\ref#1{{\goodbreak\parindent=0pt\parskip=0pt \hangindent=10pt
  \hangafter=1 #1.\smallskip\goodbreak}}
\def\title#1{{\it #1}.}

%
%

%
\def\caption#1{\vbox{\vskip12pt$${\hbox{\vbox%
{\hsize=128truemm\noindent#1}}}$$\vskip-12pt}}
%
%
\frenchspacing
\voffset=2\baselineskip
\def\maketitle{
{{\noindent\baselineskip=20pt\titlefont\title\par}\vskip20pt
{{\noindent\baselineskip=16pt\namefont\author\par}
}}}
\nopagenumbers
\hfuzz=1.5pt
\vfuzz=1.5pt
\vsize=225truemm
\hsize=138truemm
\parskip=0pt plus 1pt
\parindent=10pt
\baselineskip=12pt
%
%
\def\footline{{\baselineskip=18pt\ifnum\Headernumber=0
{\line{\centerline{\rm\folio}}}\else{}\fi\noindent}}
\output={
  \shipout\vbox{
    \ifnum\Headernumber=0
{\vbox to18pt{\hbox to\hsize{\hfill}\vfill}}
     \vbox to\vsize{\boxmaxdepth\maxdepth\pagecontents}
          {\vbox to18pt{\vfil{\hbox to\hsize{\footline}}}}
    \else{\nopagenumbers\vbox to18pt{\hbox to\hsize{\headline}\vfil}}
               \vbox to\vsize{\boxmaxdepth\maxdepth\pagecontents}
      {\vbox to18pt{\hbox to\hsize{\hfill}\vfill}}
     \fi
}
\global\advance\pageno1\global\advance\Headernumber by 1
\ifnum\outputpenalty>-1000000 \else \dosupereject\fi}
\def\Map{\mathop{\rm Map}}
\def\newp#1{\bigskip\noindent{\bf #1.}\enspace}
\def\g{\mathord{\bf g}}
\let\tto=\longrightarrow
\def\author{Jos\'e M Figueroa-O'Farrill}
\def\title{Obstructions to Gauging WZ Terms:\hfil\break
\phantom{b}a Symplectic Curiosity}\footnote{}{Talk given at the
Workshop on the Geometry of Constrained Dynamical Systems at the Isaac
Newton Institute for Mathematical Sciences, June 14-18, 1994.}
\maketitle
\def\title{Obstructions to Gauging WZ Terms: a Symplectic Curiosity}
\vskip 15truemm
\line{\hfill\sl Dedicated to the memory of Arnoldo Ferrer Andreu
(1916-1994)}
\vskip 1truecm

\newp0
As children we are taught to expect that behind any number of
continuous symmetries of a dynamical system, there always lurk an
equal number of conserved quantities.  However at some point in our
lives we find out that this is not necessarily the case.  The
correspondence between symmetries and conservation
laws---equivalently, the existence of a moment mapping associated to a
symplectic group action---must overcome a homological obstruction.
That is, this obstruction takes the form of a class in a suitably
defined cohomology theory which must vanish for the correspondence to
go through.  The purpose of this talk is to point out a curious
coincidence.  In my joint work with Sonia Stanciu trying to understand
the gauging of nonreductive Wess-Zumino-Witten models, I came across
the fact that the obstructions to gauging the Wess-Zumino term of a
(toy) one-dimensional $\sigma$-model are none other than the
obstructions for the existence of the moment mapping.  Of course, as a
physical system this $\sigma$-model is not very interesting, but I
hope that this symplectic curiosity serves to bring a little {\it
divertimento\/} to fit the occasion.

It is a pleasure to thank Chris Hull, Takashi Kimura, and Sonia
Stanciu for discussions on this topic; and especially Jim Stasheff for
comments on a previous version of the {\TeX}script.  I would also like
to express my gratitude to John Charap for organizing the conference
and giving me the opportunity to present this talk.

\newp1
Let $(M,\omega)$ be a symplectic manifold; that is, the two-form
$\omega$ is closed and is nondegenerate when thought of as a section
of $\mathop{\rm Hom}\,(TM,T^*M)$.  We say that a vector field $\xi$ on
$M$ is {\bf symplectic} if its flow fixes $\omega$:
$$
{\cal L}_\xi \omega = 0~.
$$
Since $d\omega=0$, this means that the one-form $\imath(\xi)\omega$ is
closed.  If $\imath(\xi)\omega$ is actually exact---so that there is a
function $f$ such that $\imath(\xi)\omega = df$---then $\xi$ is
called {\bf hamiltonian}.  We see then that in a symplectic manifold
one has the following interpretation of the first de~Rham cohomology:
$$
H^1(M) = {\hbox{closed one-forms}\over\hbox{exact one-forms}}
       = {\hbox{symplectic vector fields}\over\hbox{hamiltonian vector
fields}}~.
$$
In other words, we have an exact sequence of vector spaces
$$
0 \tto \mathord{\rm Ham}(M) \tto \mathord{\rm Sym}(M) \tto H^1(M) \tto
0~,
\eqno(1.1)
$$
where $\mathord{\rm Ham}(M)$ and $\mathord{\rm Sym}(M)$ denote the
hamiltonian and symplectic vector fields, respectively.  It is clear
from its definition as the stabilizer of $\omega$, that $\mathord{\rm
Sym}(M)$ is a Lie algebra.  Moreover, $\mathord{\rm Ham}(M)$ is a Lie
subalgebra.  Indeed, if $\xi_f$ and $\xi_g$ are hamiltonian vector
fields associated to the functions $f$ and $g$,
then
$$
[\xi_f, \xi_g] = \xi_{\{f,g\}}\eqno(1.2)
$$
where $\{f,g\}$ is the Poisson bracket.  More is true, however, and
$\mathord{\rm Ham}(M)$ is actually an ideal of $\mathord{\rm Sym}(M)$;
for if $\eta$ is a symplectic vector field
$$
[\eta,\xi_f] = \xi_{\eta\cdot f}~.
$$
In other words, the exact sequence (1.1) is actually an exact sequence
of Lie algebras.  The induced Lie bracket on $H^1(M)$ is zero,
however, because of the fact that $\mathord{\rm Ham}(M)$ contains the
first derived ideal $\mathord{\rm Sym}(M)' \equiv [\mathord{\rm Sym}(M),
\mathord{\rm Sym}(M)]$.

The assignment of a hamiltonian vector field to a function defines a
map
$$
\eqalign{C^\infty(M) &\to \mathord{\rm Ham}(M)\cr
         f &\mapsto \omega^{-1}(df)\cr}
$$
which by (1.2) is a Lie algebra morphism.  Its kernel consists of
the locally constant functions $df=0$; that is, $H^0(M)$.  This gives
rise to another exact sequence of Lie algebras
$$
0 \tto H^0(M) \tto C^\infty(M) \tto \mathord{\rm Ham}(M) \tto
0~,\eqno(1.3)
$$
where $H^0(M)$ is the center of $C^\infty(M)$ and hence abelian.
Putting this sequence together with (1.1) we find the following 4-term
exact sequence of Lie algebras interpolating between $H^0(M)$ and
$H^1(M)$:
$$
0 \tto H^0(M) \tto C^\infty(M) \tto \mathord{\rm Sym}(M) \tto H^1(M)
\tto 0~.\eqno(1.4)
$$

\newp2
Now let $G$ be a connected Lie group acting on $M$ in such a way that
$\omega$ is $G$-invariant.  Let $\g$ denote the Lie algebra of
$G$.  Every $X\in\g$ gives rise to a Killing vector field on $M$ which
we denote $\xi_X$.  The map $X\mapsto \xi_X$ is a Lie algebra
morphism.  Since $\omega$ is $G$-invariant, $\xi_X$ is symplectic.
In other words, a symplectic $G$-action on $M$ gives rise to a Lie
algebra morphism $\g \tto \mathord{\rm Sym}(M)$.  There will be
conserved charges associated to these continuous symmetries if and
only if this map lifts to a Lie algebra morphism $\g\tto
C^\infty(M)$ in such a way that the resulting diagram
$$
\matrix{0 &\to& H^0(M) &\to& C^\infty(M)& \to &\mathord{\rm
Sym}(M)& \to &H^1(M)& \to& 0\cr
&&&&\nwarrow&&\nearrow&&&&\cr
&&&&&\g&&&&&\cr}
$$
commutes.  The obstruction to the existence of such a lift follow
easily from the exactness of (1.4).  First of all, the image of $\g$
in $\mathord{\rm Sym}(M)$ will come from $C^\infty(M)$ if it is sent
to zero in $H^1(M)$.  That is, if there exists functions $\phi_X$ such
that $\imath(\xi_X)\omega = d\phi_X$.  This is not enough because we
want the map $X\mapsto\phi_X$ to be a Lie algebra morphism.  Because
the map $X\mapsto \xi_X$ is a Lie algebra morphism, the map
$X\mapsto\phi_X$ is at most a projective representation characterized
by the $H^0(M)$-valued cocycle $c(X,Y) \equiv \{\phi_X,\phi_Y\} -
\phi_{[X,Y]}$.  If and only if this cocycle is a coboundary is the
representation an honest representation.  Indeed, if there exists some
map $X\mapsto b_X\in H^0(M)$ such that $c(X,Y) = - b_{[X,Y]}$, then
one straightens the map $X\mapsto \phi'_X = \phi_X - b_X$ and the
resulting map $\g \to C^\infty(M)$ is a morphism.

If this is case then one can define the {\bf moment(um) mapping}
$$
\Phi:M\tto \g^*
$$
by $\langle \Phi(m),X\rangle = \phi'_X(m)$ for all $m\in M$.  This map
is equivariant in that it intertwines between the $G$-action on $M$
and the coadjoint action on $\g^*$.

\newp3
We can understand the conditions
$$
\eqalignno{&\imath(\xi_X)\omega = d\phi_X &(3.1a)\cr
           &\{\phi_X,\phi_Y\} = \phi_{[X,Y]} &(3.1b)\cr}
$$
purely in terms of cohomology as follows.  First of all notice that
the map $\g \to H^1(M)$ defined by $X\mapsto [\imath(\xi_X)\omega]$
annihilates the first derived ideal $\g'$, since $[\mathord{\rm
Sym}(M), \mathord{\rm Sym}(M)]\subset \mathord{\rm Ham}(M)$.
Therefore it induces a map $\g/\g' \to H^1(M)$; or, in other words, it
defines an element in
$$
\left(\g/\g'\right)^* \otimes H^1(M) \cong H^1(\g)\otimes H^1(M)~.
$$
Then (3.1a) simply says that this element is zero.  Similarly the
cocycle $c:\bigwedge^2\g \to H^0(M)$ defined above defines a class in
$H^2(\g)\otimes H^0(M)$.  Then (3.1b) says that this class should be
zero.  In other words, the obstruction to the existence of a moment
mapping defines a class
$$
[O] \in \left( H^1(\g)\otimes H^1(M)\right) \oplus \left( H^2(\g)
\otimes H^0(M) \right)~.\eqno(3.2)
$$

In fact, we can understand this class as a single class in a different
cohomology theory.  Let us start by considering the $G$-action on $M$ as
a map
$$
\alpha: G\times M \tto M
$$
and let us define a $G$-action on $G\times M$ to make $\alpha$
equivariant.  One convenient way to do so is
$$
\beta: G \times G\times M \tto G \times M
$$
where $\beta(g,h,m) = (gh,m)$; that is, $G$ acts via left
translations on the first factor and ignores the second.  Equivariance
of $\alpha$ allows us to pull back $G$-invariant forms on $M$ to
$G$-invariant forms on $G\times M$.  The $G$-invariant forms on
$G\times M$ form a subcomplex $\Omega^{\cdot}(G\times M)^G$ of the
de~Rham complex.  Therefore $\alpha^*\omega \in \Omega^2(G\times
M)^G$.  Similarly if we denote by $\pi : G\times M \to M$ the
Cartesian projection onto the second factor, $\pi^*\omega$ is also a
$G$-invariant form on $G\times M$.  Define then $\omega_\alpha \equiv
\alpha^*\omega - \pi^*\omega$.  This is a closed form in
$\Omega^2(G\times M)^G$ and hence defines a class in $H^2(G\times
M)^G$.  The complex $\Omega^{\cdot}(G\times M)^G$ is isomorphic to the
double complex $\Omega^{\cdot}(G)^G\otimes \Omega^{\cdot}(M)$.
Applying the K\"unneth theorem to this complex, one finds that
$$
H^n(G\times M)^G \cong \bigoplus_{p+q=n} H^p(\g)\otimes
H^q(M)~.\eqno(3.3)
$$
It is then an easy computational matter to prove that under this
isomorphism the class of $\omega_\alpha$ goes over to the class $[O]$
in (3.2).  (The $H^0(\g)\otimes H^2(M)$ component is zero precisely
because in $\omega_\alpha$ we subtract $\pi^*\omega$ from
$\alpha^*\omega$.)

As an example, if $(T^*N, d\theta)$ is the phase space of some
configuration space $N$ on which $G$ acts, the action of $G$
lifts naturally to a symplectic action on $M$.  In fact, the
tautological one-form $\theta$ is already invariant.  In this case,
$\omega_\alpha = d(\alpha^*\theta - \pi^*\theta)$ and since
$(\alpha^*\theta - \pi^*\theta)$ is $G$-invariant, the class
$[\omega_\alpha]$ in $H^2(G\times M)^G$ is trivial.  Our ``classical''
intuition on the correspondence between continuous symmetries and
conservation laws is borne out of this example.

\newp4
What does this have to do with gauging $\sigma$-models? Let $B$ be a
two-manifold with boundary $\partial B = \Sigma$.  Let $(M,\omega)$ be
as before except that we drop the nondegeneracy condition on $\omega$.
The Wess-Zumino term of the $\sigma$-model in question is given by the
function
$$
S_{\rm WZ}[\varphi] = \int_B \varphi^* \omega\eqno(4.1)
$$
on the space of maps $\varphi: B \to M$; but because $\omega$ is
closed, the resulting equations of motion only depend on the
restriction of $\varphi$ to the boundary $\Sigma$.  Therefore it
defines a variational problem in the space $\Map(\Sigma,M)$ of maps
$\Sigma\to M$ (which extend to $B$).  The $\sigma$-model also comes
with a kinetic term defined on $\Sigma$, but since the gauging of this
term is simply accomplished via minimal coupling we shall disregard it
in what follows.  It should also be mentioned that we are ignoring for
the present purposes the topological obstructions concerning the
well-definedness of the WZ term itself.  Similarly we will consider
only gauging the algebra: demanding invariance under ``large'' gauge
transformations invariably brings about other topological
obstructions.

Let $G$ be a connected Lie group, acting on $M$ in such a way that it
fixes $\omega$.  The action of $G$ on $M$ induces an action of $G$ on
$\Map(B,M)$ under which the action (4.1) is invariant.  For our
purposes, gauging the WZ term will consist in promoting (4.1) to an
action which is invariant under $\Map(\Sigma,\g)$ via the addition of
further terms involving a gauge field.  We do this in steps following
the Noether procedure.

\newp5
Let $\lambda\in\Map(\Sigma,\g)$.  More explicitly, if we fix a basis
$\{X_a\}$ for $\g$, then $\lambda = \lambda^a X_a$ with $\lambda_a$
functions on $\Sigma$.  The action of $\lambda$ on the pull-back of
any form $\Omega$ on $M$, is given by
$$
\delta_\lambda \varphi^*\Omega = d\lambda^a\wedge
\varphi^*\imath_a\Omega + \lambda^a \varphi^*{\cal L}_a\Omega
$$
where $\imath_a$ and ${\cal L}_a$ denote respectively the contraction
and Lie derivative relative to the Killing vector corresponding to
$X_a$.  In particular since $\omega$ is closed and $\g$-invariant, we
find that $\delta_\lambda \varphi^*\omega = d\left(\lambda^a \varphi^*
\imath_a\omega\right)$, whence the variation of (3.1) becomes
$$
\delta_\lambda S_{\rm WZ}[\varphi] = \int_\Sigma \lambda^a\,
\varphi^*\imath_a\omega~.
$$
Let us now introduce a $\g$-valued gauge field $A = A^a X_a$ on
$\Sigma$, which transforms under $\Map(\Sigma,\g)$ as
$$
\delta_\lambda A = d\lambda + [A,\lambda]~.
$$
The most general (polynomial) term we can add to (3.1) involving the
gauge field is given by
$$
S_{\rm extra}[\varphi,A] = \int_\Sigma A^a\,\varphi^* \phi_a
$$
for some functions $\phi_a \in C^\infty(M)$.  It is then a small
computational matter to work out the conditions under which the total
action
$$
S_{\rm GWZ}[\varphi,A] = \int_B \varphi^* \omega + \int_\Sigma
A^a\,\varphi^* \phi_a
$$
is gauge-invariant; that is, $\delta_\lambda S_{\rm GWZ} = 0$.  Doing
so one finds that the conditions are
$$
\eqalign{&\imath_a\omega = d\phi_a \cr
           &{\cal L}_a \phi_b = {f_{ab}}^c\phi_c \cr}
$$
which are none other than (3.1a) and (3.1b) relative to the chosen
basis for $\g$.

We therefore conclude that, for $\omega$ a symplectic form, the WZ
term (4.1) can be gauged if and only if one can define an equivariant
moment mapping for the $G$-action.

\medskip
\noindent{\bf 6.}\enspace{\bf Bibliography}
\smallskip\noindent
The homological obstructions to defining a moment mapping have been
well-known since at least the mid nineteen-seventies.  The treatment
here follows in spirit the one in Weinstein's 1976 lectures [We].  The
conditions for gauging reductive (two-dimensional) WZW models were
obtained independently by Hull and Spence in [HS1] and by Jack, Jones,
Mohammedi and Osborne in [JJMO]. The conditions for gauging
higher-dimensional $\sigma$-models with WZ term were later considered
by Hull and Spence in [HS2].  The conditions for gauging nonreductive
WZW models have been obtained in [FS1] as part of a general analysis
of such models.  The homological (re)interpretation of the
obstructions to gauging a general WZ term will appear in [FS2], where
they are understood in terms of the equivariant cohomology.

\medskip

\def\NPB#1#2#3{{\sl Nucl. Phys.} {\bf B#1} (#2) #3}
\def\PLB#1#2#3{{\sl Phys. Lett.} {\bf #1B} (#2) #3}
\item{[FS1]}{JM Figueroa-O'Farrill and S Stanciu, {\sl Nonreductive WZW
models\/}, in preparation.}
\item{[FS2]}{JM Figueroa-O'Farrill and S Stanciu, {\sl Equivariant
cohomology and gauged bosonic $\sigma$-models\/}, {\tt
hep-th/9407149}.}
\item{[HS1]}{CM Hull and B Spence, \PLB{232}{1989}{204}.}
\item{[HS2]}{CM Hull and B Spence, \NPB{353}{1991}{379-426}.}
\item{[JJMO]}{I Jack, DR Jones, N Mohammedi, and H Osborne,
\NPB{332}{1990}{359}.}
\item{[We]}{A Weinstein, {\sl Lectures on Symplectic Manifolds\/},
CBMS Regional Conference Series on Mathematics, {\bf 29}, 1977.}

\newp7
{\bf Postscript}
\smallskip\noindent
After the talk, J.~Cari\~nena pointed out to me another way to
understand the obstructions in (3.1) in terms of Lie algebra
cohomology with coefficients in the exact one-forms (equivalently, the
hamiltonian vector fields).  If we think of (1.1) and (1.3) as exact
sequences of $\g$-modules, we obtain two long exact sequence in Lie
algebra cohomology.  The map $X\mapsto \xi_X$ defines a class in
$H^1(\g; \mathord{\rm Sym}(M))$.  By exactness of the sequence induced
by (1.1), we see that it comes from $H^1(\g; \mathord{\rm Ham}(M))$ if
and only if its image in $H^1(\g; H^1(M))$ vanishes.  Supposing it
does and using now the exactness of the sequence induced by (1.3), we
see that this class in $H^1(\g; \mathord{\rm Ham}(M))$ comes from
$H^1(\g; C^\infty(M))$ precisely when its image in $H^2(\g;H^0(M))$
vanishes.  These two obstructions precisely correspond to the classes
in (3.2).  Finally, I was informed by G.~Papadopoulos, that the
obstructions in (3.2) can also be understood as ``anomalies'' to
global symmetries in the quantization of a particle interacting with a
magnetic field.  The details appear in {\sl Comm.~Math.~Phys.\/} {\bf
144} (1992) 491-508.  I am grateful to them both for letting me know
of their results during the conference.

\bye